\documentclass[a4paper,twocolumn] {article} 
\pdfoutput=1 
\usepackage{jinstpub} 
\usepackage{graphicx}
\usepackage{subfigure}
\usepackage{hyperref}

\title{\boldmath Integrated cooling channels in position-sensitive silicon detectors}

\newcommand{\Ohm}{\ensuremath{\Omega}}
\newcommand*{\degr}{\ensuremath{^\circ}}

\author[a]{L.~Andricek}
\author[b]{M.~Boronat}
\author[b]{J.~Fuster}
\author[b,1]{I.~Garcia\note{Corresponding author.}}
\author[b]{P.~Gomis}
\author[c]{C.~Marinas}
\author[a]{J.~Ninkovic}
\author[b]{M.~Perell\'{o}}
\author[b]{M.A.~Villarejo}
\author[b]{M.~Vos}

\affiliation[a]{HalbLeiterLabor, Max Plank Gesellschaft, Munich, Germany}
\affiliation[b]{IFIC (UV-CSIC) Valencia, Spain}
\affiliation[c]{University of Bonn, Germany}

\emailAdd{ignacio.garcia@ific.uv.es}

\abstract{We present an approach to construct position-sensitive silicon detectors with an integrated cooling circuit. Tests on samples demonstrate that a very modest liquid flow very effectively cool the devices up to a power dissipation of over 10~W/cm$^2$. The liquid flow is found to have a negligible impact on the mechanical performance. A finite-element simulation predicts the cooling performance to an accuracy of approximately 10~\%.}

\arxivnumber{} 
\begin{document}
\maketitle
\flushbottom

\section{Introduction}
\label{Introduction}
Micro-Channel Cooling is used extensively in industry to cool devices
with a large power density that are to be employed in areas where the material
involved in removing the power must be reduced to the minimum. An obvious
application in particle physics  
are the pixel detectors used in collider experiments to reconstruct the 
creation vertices and trajectories of charged particles. The requirement
of highly granular detectors (and thus large channel density) and fast read-out
(requiring larger power in the Front-End) lead to typical power densities of
several 100 $mW/cm^2$. At the same time the detector material - and that in 
supports and services in the detector acceptance - must be reduced to the bare 
minimum to avoid the effect of multiple scattering on the track parameter 
resolution. Detector concepts~\cite{bib:ilddbd} for future electron-positron 
colliders~\cite{Baer:2013cma,Behnke:2013lya,Linssen:2012hp,Gomez-Ceballos:2013zzn}
aim for a total material budget of 0.12-0.2\% $X_0$/layer, where $X_0$ 
is the radiation length. This is equivalent to only 100-200 $\mu\mathrm{m}$ of
silicon. State-of-the art sensors can be thinned to well below this. 
One of main challenges is therefore to reduce the material involved in services
and supports to below the 0.1\% $X_0$ level.

Micro-cooling channels integrated in the detector itself provide a very 
effective means of removing heat. Bringing the cooling circuit to
within hundreds of microns of the heat source and removing thermal
barriers (material interfaces, glue layers) reduces the temperature gradient
between the heat source and the cooling liquid. The minimal extra material
in cooling fluid and circuits helps to reduce the material involved in
removing the heat from the detector. 
Compared to other low-mass proposals (in particular cooling via an air 
flow in the detector volume) micro-channel cooling is expected to provide 
a much tighter control over the local temperature.

Several groups have produced micro-cooling circuits in (non-active) 
silicon wafers for application in high energy physics experiments~\cite{Mapelli:2012zz,Abelevetal:2014dna,Buytaert:2013hka,Nussle:2014sza}. Micro-channel
cooling is part of the detector concepts for LHCb, NA62 and ALICE. In all
these concepts the micro-channel circuit is implemented in a low-mass
silicon cooling plate that is brought into thermal contact with the
detector that is to be cooled, typically through a thin glue layer.

In this paper we take the integration of
cooling and active detector elements one step further by monolithically
integrating the micro-channel cooling circuit in a silicon sensor. 
Our approach fits in a low-pressure, low-mass, mono-phase cooling solution
for applications with a relatively modest power density and room-temperature
operation, in combination with very strict requirements on the material budget.  
We validate a production process that combines commercial micro-electronics
processing with wafer thinning and the etching of a micro-cooling circuit.
We develop a custom connector and a finite element simulation to allow
rapid feed-back from laboratory measurement and simulation. Finally, we 
characterize the thermo-mechanical performance of prototypes.

In Section~\ref{sec:process} we present an extension of the process that is 
used to produce the thin, self-supporting silicon DEPFET sensors for 
Belle-II~\cite{Abe:2010gxa} and the ILC~\cite{Alonso:2012ss}, with an extra step to etch a 
cooling manifold into the wafer. The custom interface of the micro-channel
cooling circuit to commercial high-pressure laboratory equipment
is presented in Section~\ref{sec:connectors}. Section~\ref{sec:simulation}
introduces a finite element simulation of the micro-channel circuit
and Section~\ref{sec:laboratory} gives an overview of the setup
in the laboratory. 
We present results from a characterization of the local cooling 
performance of prototypes in Section~\ref{sec:results}, comparing
the laboratory results to a finite-element simulation.
The impact on the mechanical stability is reported in 
Section~\ref{sec:results3}. 
Finally, the conclusions are
presented in Section~\ref{sec:conclusions}.

\section{Integration of the cooling circuit}
\label{sec:process}

The approach we follow is based on the silicon-on-insolator thinning 
concept~\cite{VTXdepfetmech2} developed for DEPFET active pixel 
detectors, the technology chosen for the Belle II vertex 
detector~\cite{Abe:2010gxa} and a candidate for the linear collider
experiments~\cite{Alonso:2012ss}. 
The process uses two silicon wafers: the top (sensor) 
wafer forms the active detector material, while the bottom (handle) wafer 
forms the supporting frame of the all-silicon ladder and, in this 
implementation, hosts the micro-channel cooling circuit. 

\begin{figure*}[t!]
{
\centering
{
\includegraphics[width=0.95\textwidth]{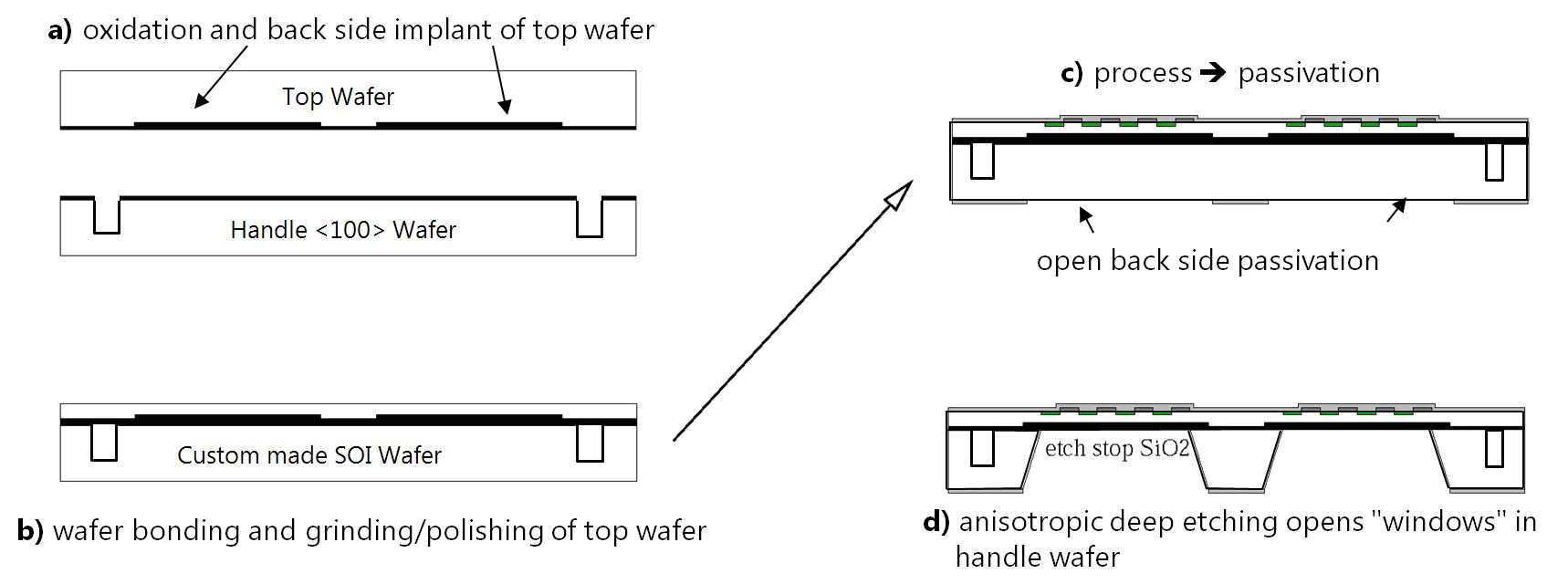}}

\caption{The process sequence for production of thin silicon sensors 
with electrically active back side implant and integrated cooling channels 
starts with the oxidation of the top and handle wafer and the back side 
implantation for the sensor devices; cooling channels are etched into 
the handle wafer before bonding (a). After direct wafer bonding, the top 
wafer is thinned and polished to the desired thickness (b). The processing 
of the devices on the top side of the wafer stack is done on conventional 
equipment; the openings in the back side passivation define the areas 
where the bulk of the handle wafer will be removed (c). The bulk of 
the handle wafer is removed by deep anisotropic wet etching. The etch 
process stops at the silicon oxide interface between the two wafers (d). 
The cooling channels are only accessible after dicing of the wafers.
}
\label{fig:process}	
}
\end{figure*}

The process is schematically represented in Fig.~\ref{fig:process} and
can be summarized as follows:
\begin{itemize}
\item As the first step both top and handle wafer are thermally oxidized 
yielding a SiO$_2$ isolation of several 100nm. For sensors requiring an active 
backside, the backside implantation is done after an optional 
photolithography step for the definition of the implanted areas.
\item A photolithography step on the handle wafer defines the 
micro-channels. The SiO$_2$ layer is etched acting as a hard mask for the 
etching dry reactive ion etching (DRIE) of the handle wafer silicon. 
This DRIE step yields the micro-channels. The typical depth of the 
channels is a few 100 $\mu\mathrm{m}$, determined by the etching time of the 
DRIE process. In the present case the result of the process are 
rectangular channels with a depth of about 380 $\mu\mathrm{m}$ and 
varying width. After this step, the photoresist and the SiO$_2$ layer 
are removed. 
\item After etching the micro-manifold 
the top and the handle wafer are joined utilizing 
a direct wafer bonding process. The two wafers are initially held by 
van-der-Waals forces.  After a high-temperature step beyond 1000\degr the 
bond between the wafers is transformed to Si-O-Si bonds. The two wafer 
wafers are electrically separated by the SiO$_2$ of the top wafer. The top 
wafer is now thinned down by grinding and a chemical-mechanical 
polishing (CMP) process to the desired 
thickness of the sensor material. While this is basically a free parameter, 
practical considerations (signal-to-noise ratio, mechanical stability) 
suggest a thickness between 30 $\mu\mathrm{m}$ and 75 $\mu\mathrm{m}$.
\item The SOI wafers with integrated cavities in the handle wafer 
(CSOI) can be processed on a standard semiconductor processing line. 
In this step the sensor part on the top wafer is finished including the 
metal layers yielding a fully functional thin sensor material directly 
bonded to the handle wafer.
\item Finally, at the end of the process sequence the oxide at 
the backside of the handle wafer is opened where the support material 
is not required. In a wet chemical etching step, the silicon is removed 
in a Tetramethyl Ammonium Hydroxid (TMAH) solution. The etch process stops 
at the buried SiO$_2$ layer between handle and top wafer. After cutting 
the buried micro-channels are exposed and ready to be connected to the 
cooling circuit.  
\end{itemize}

This process is fully compatible with use of the silicon as the active
material in a position-sensitive device. The sensor wafer can be fully
depleted through application of a bias voltage. All steps except the 
insertion of the micro-manifold are identical to the DEPFET manufacturing 
process for Belle II.

\section{Microchannel circuits}
\label{sec:microchannels}

\begin{figure}[h!]
\centering
\includegraphics[width=0.9\columnwidth]{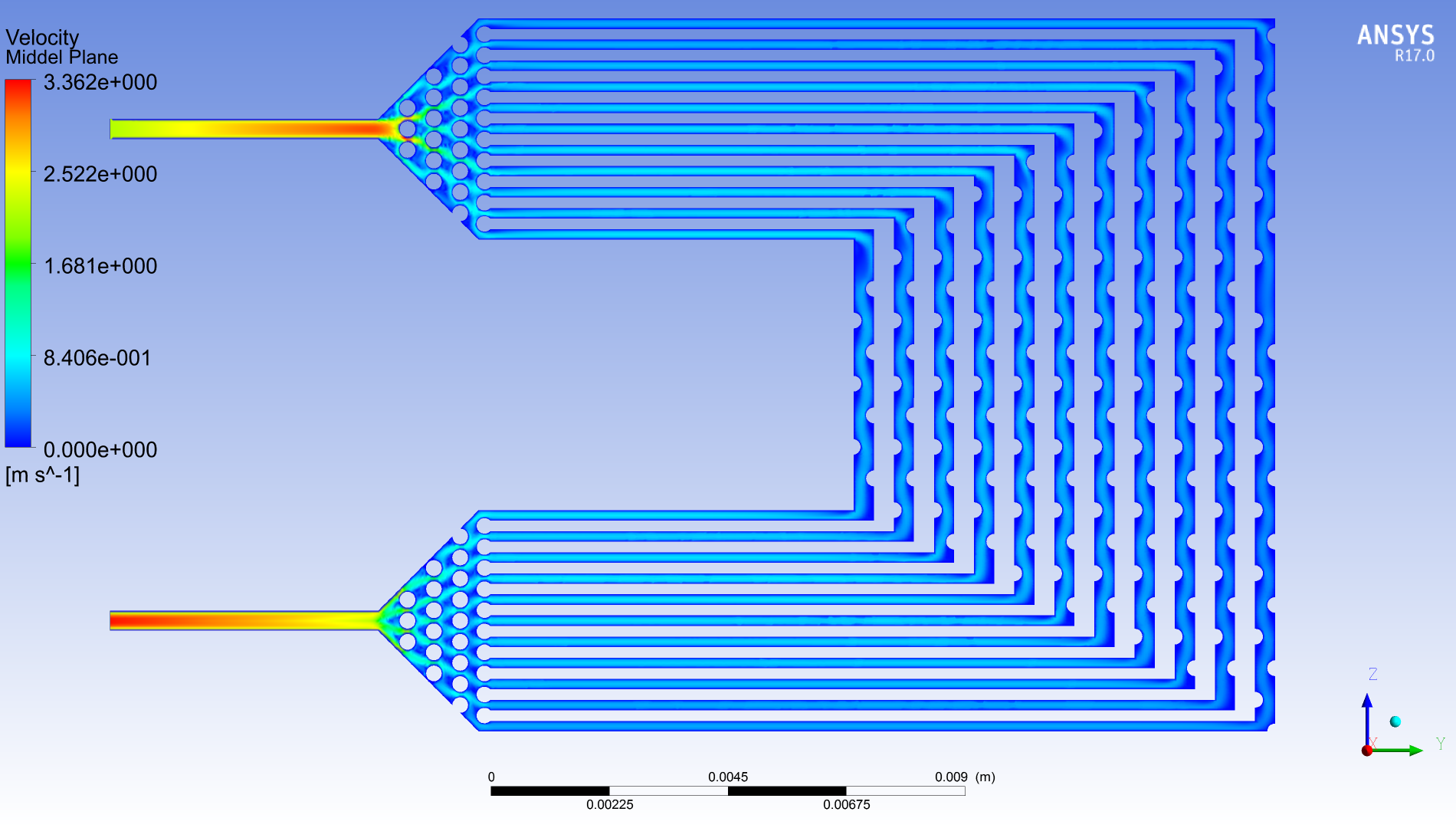}
\caption{Design of the micro-channel manifold. The colour code represents the expected velocities in the center of the cooling channels, for different sections of the manifold for a total flow rate of 1 l/h. }
\label{fig:manifold}
\end{figure}

To evaluate the cooling performance of a micro-cooling circuit integrated in the sensor we produced a number of mechanical samples. The design is shown in Figure~\ref{fig:manifold}. The inlet and outlet are shown on the leftmost side of the design. Both have a length of 5 mm and an approximately rectangular cross-section of 340 $\times$ 380 $\mu\mathrm{m}^2$. A manifold with multiple cooling lines in parallel is formed by fanning out the inlet in the $\Delta$ area. The single channel is divied into 11 separate channels. Circular {\em pillar} structures are present in the $\Delta$ area to support the thin silicon layer spanning a distance of up to 500~$\mu\mathrm{m}^2$.  

\begin{figure*}[t!]
\centering
\subfigure[]{
\includegraphics[width=0.52\columnwidth]{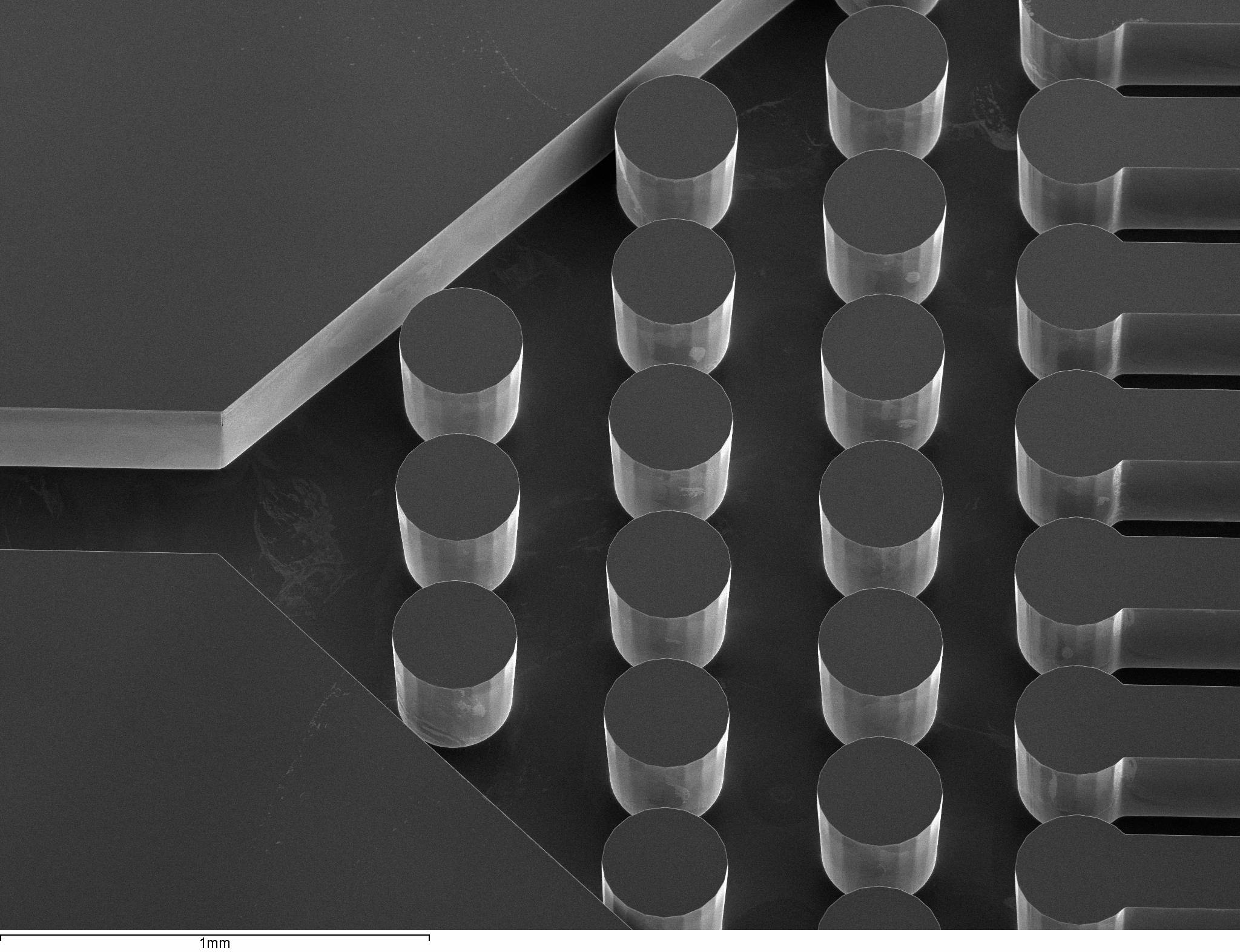}
}
\subfigure[]{
\includegraphics[width=0.52\columnwidth]{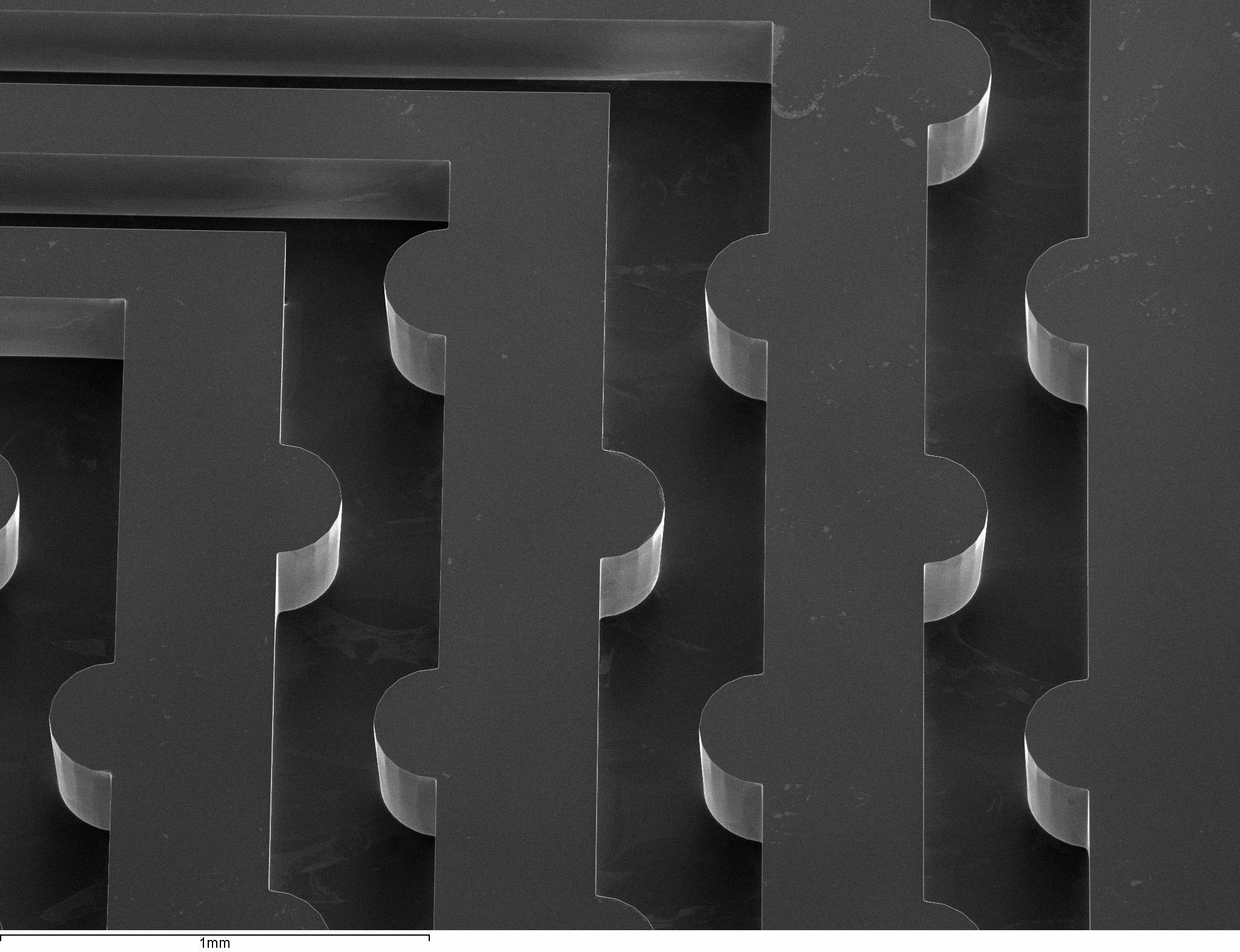}
}
\subfigure[]{
\includegraphics[width=0.4\columnwidth, angle=90]{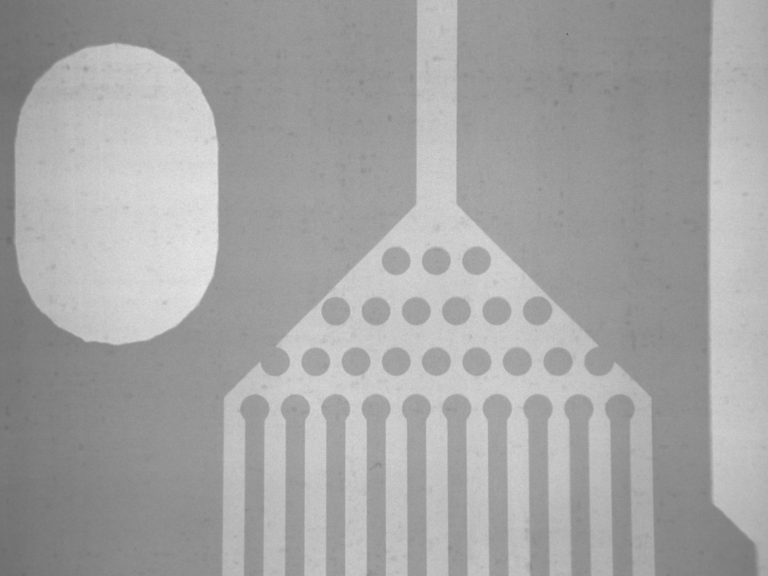}
}
\subfigure[]{
\includegraphics[width=0.4\columnwidth, angle=90]{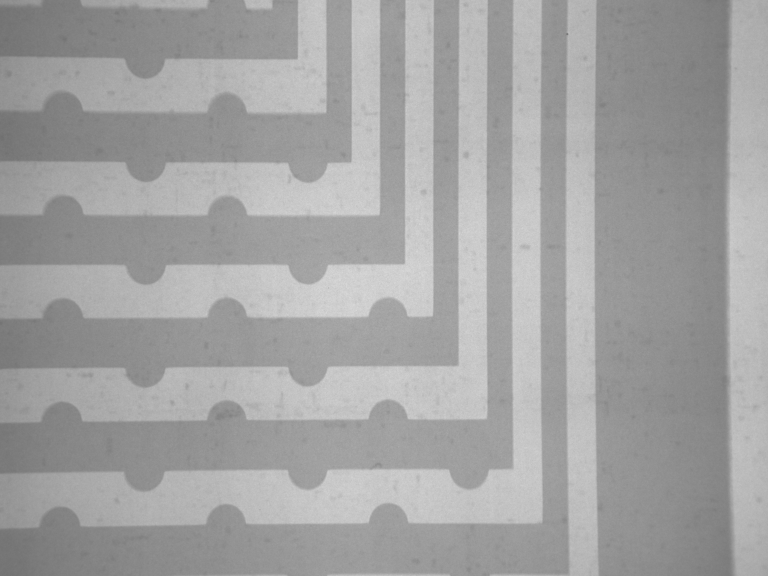}
}
\caption{Photographs of the micro-channel cooling manifold etched into the {\em support} wafer before bonding the {\em sensor} wafer that closes the circuit. The photograph (a) zooms in on the position where the inlet (on the left of the image) fans out into several cooling channels. The circular structures are pillars that support the thin layers of silicon covering the triangular area. The photograph (b) shows a detail of several of the cooling channel. X-ray images of the same structures in the micro-channel manifolds after sealing the circuit with the top ({\em sensor}) wafer are shown in (c) and (d).}
\label{fig:manifold_etching}
\end{figure*}

The 11 cooling channels have a cross-section of 200 $\times$ 340 $\mu\mathrm{m}^2$ (width $\times$ height). The lengths of the channels vary according to their location; the length of the inner loop is 20~mm, that of the outer loop 40~mm. The expected liquid flow in the manifold is indicated with colours in Figure~\ref{fig:manifold} (the Finite Element simulation used to obtain this information is discussed in Section~\ref{sec:simulation}). Clearly, the liquid velocity is highest in the inlet and outlet pipes.

The manifolds are etched into a 450-525 $\mu\mathrm{m}$ thick silicon 
handle wafer. 
The photographs of Figure~\ref{fig:manifold_etching} present two details of the circuit. The photograph on the left zooms in on the position where the inlet (on the left of the image) fans out into several cooling channels. The circular structures are pillars that support the thin layers of silicon covering the triangular area. The photograph on the right shows a detail of several of the corner of four cooling channels. The X-ray images in Figure~\ref{fig:manifold_etching} (c) and (d) present detailed views of the same areas of the circuit after sealing the circuit with the top wafer.

\begin{figure*}[t!]
\centering
\includegraphics[width=0.9\textwidth]{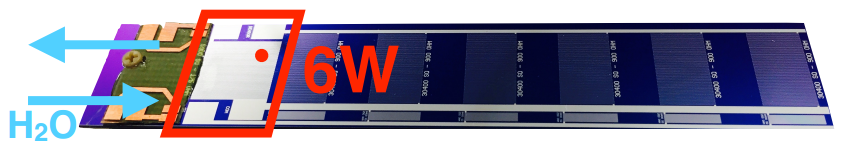}
\caption{A 540 $\mu\mathrm{m}$ thick silicon ladder with an integrated micro-channel cooling circuit. The micro-manifold is located in the end-of-ladder area, indicated with a red box in the image. The arrows indicate indicate the approximate locations of the inlet and outlet channels. The resistive circuit are visible as grey areas. }
\label{fig:dummy}
\end{figure*}

To tackle the connectivity problems and assess the thermo-mechanical performance a small number of test structures was designed and produced for a 
characterization in the laboratory. The production process follows closely the 
steps outlined in Section~\ref{sec:process}, but simplifies the steps not
directly related to the thermal performance. The back-side processing
mentioned in the first step is omitted. The top-side processing is 
limited to the implementation of a single metal layer. For these first samples
the handle wafer thinning step is not performed\footnote{These
steps has been performed successfully for a large number of prototypes
and are part of the production of the Belle~II vertex detector that is in 
progress at the time of writing.}. Figure~\ref{fig:dummy} 
presents a photograph of these mechanical samples.

 These thermo-mechanical samples are equipped with resistive circuits 
implemented in an Aluminium layer deposited on the surface, that 
mimic the heat-load from the electronics. Narrow traces are laid out
in a serpentine geometry to maximize the length of the trace and
thus achieve a resistance of 10-100~\Ohm. Three independent circuits
roughly correspond to the end-of-ladder electronics 
of the DEPFET ladder design for Belle~II (this is where the
read-out chips are located; the instantaneous power dissipation is
up to 6~W on less than 1 cm$^2$ of the sensor area, as indicated 
in Fig.~\ref{fig:dummy}), the balcony 
with six steering chips (running along the lower edge of the image) 
and the sensor itself (the six grey areas that cover the full width
of the sensor). The cooling manifold is located exactly under the
end-of-ladder heater circuit.

\section{Connectors}
\label{sec:connectors}
\begin{figure}[h!]
\centering
\includegraphics[width=0.9\columnwidth]{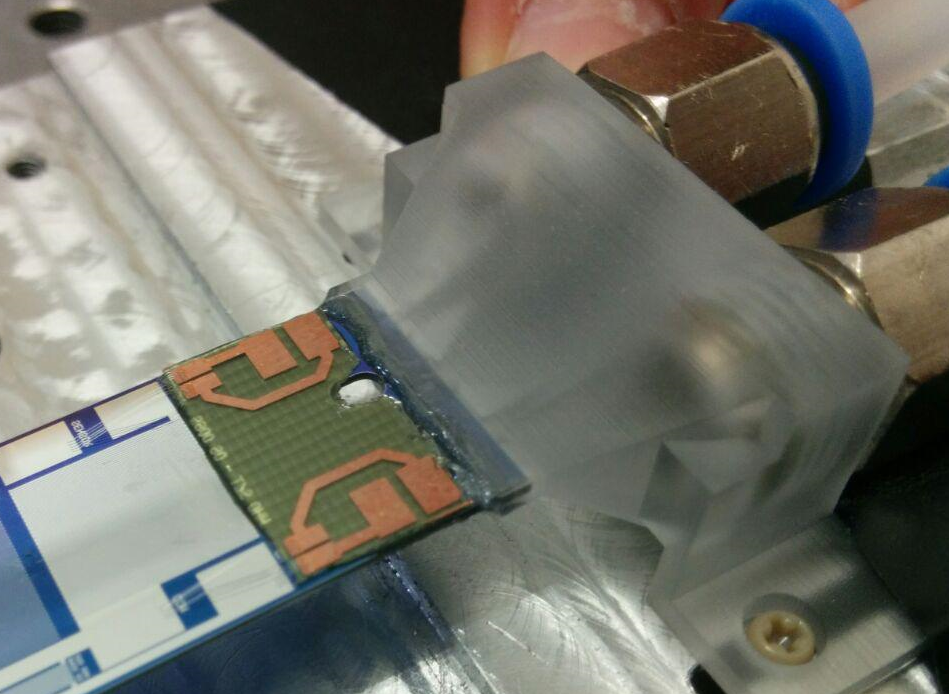} \\
\includegraphics[width=0.9\columnwidth]{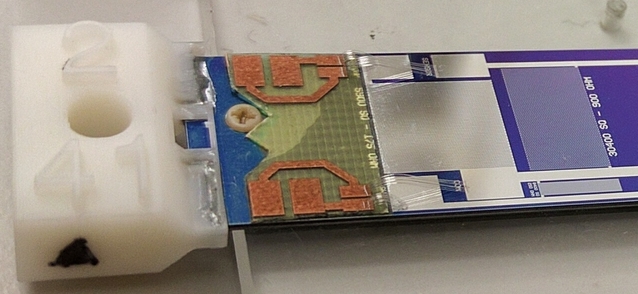}
\caption{Photographs of the silicon ladder with the integrated cooling circuit and the 3D-printed connector to the cooling circuit. Both interface the micro-channel cooling circuit to commercial fittings (they are visible in the upper right corner of the upper image). }
\label{fig:connector}   
\end{figure}

To connect the micro-channel manifold to a laboratory cooling circuit a custom 
interface is used that connects the inlet and outlet to standard commercial
high-pressure connectors. This connector consists of plastic piece that 
slides over the silicon sensor, as shown in Fig.~\ref{fig:connector}. The 
silicon joint is sealed with a glue layer (Araldite2020). On the opposite
of the connector there are two threaded holes to connect the standard fittings of the 6 mm outer diameter tubes, visible in the upper right corner of Fig.~\ref{fig:connector}. The channel in the connector gradually reduces the circular cross section of the tube fittings to a square cross section of 400 $\times$ 400 $\mu\mathrm{m}$. A photograph through the conical hole is shown in Figure~\ref{fig:connector}.
	
The connectors are produced in a 3D printer using the Stereolithography (SLO)
process offered by Sicnova. This technique offers a mechanical precision 
of 15-30 $\mu$m (depending of the 3D printing speed) and with thin layers 
of less than 300 $\mu$m. 
The material {\em visijet FX clear} with a flexural (tensile) strength of
52 MPa (82 MPa). For such plastic materials the radiation length is
typically 35-45 cm.

An assembly consisting of a connector glued onto a silicon sensor was tested 
for leaks in the CERN micro-channel cooling laboratory. No detectable increase
of the Helium level is observed, which allows to put an upper bound on the
leak rate. The maximum pressure that the connector stands is 180 bar.

The low-$Z$ material of the connector has a radiation length of approximately 40~cm. The full thickness of the connectors of the upper panel of Fig.~\ref{fig:connector} corresponds to 0.8\% $X_{0}$ for a particle crossing under a 90~\degr angle, clearly prohibitive for applications inside the tracking volume of a particle physics experiment. A miniaturized version of the connector has been designed, that offers a reduction of the material by a factor four, while maintaining compatibility with a commercial fitting standard. The result is shown in the lower panel of Figure~\ref{fig:connector}. Even smaller connectors are under development that glue a PEEK micro-tube with a diameter matching that of the micro-channel. These offer a solution with a full thickness equivalent to 0.05\% $X_0$ and may be interesting for applications that require the connectors to be in the acceptance. The development of a standard connector for micro-channel cooling is part 
of Work Package 9 of the AIDA2020 project.


\section{Finite element simulation}
\label{sec:simulation}
A finite-element simulation is used to predict the performance of different designs. The geometry presented in Section~\ref{sec:microchannels} is implemented in the CFD software Ansys 17.0. The meshed geometry has a total of 4 millions elements, which allow to observe local effects in detail.
The cooling fluid in the simulation is pure water and the solid element is silicon. The physical properties of both materials assumed in the simultion are shown in Table~\ref{tab:water}.
The boundary conditions in the simulations - the total power applied to the microchannels and the mass flow - are given in the first two columns of Table~\ref{tab:water}. They are as close as possible to the laboratory set-up. The environment temperature is set to 25\degr, similar to the ambient temperature in the laboratory. The convective heat transfer coefficient used between the sensor and the air is 5 $h=5W/m^{2}K$. The temperature of the liquid at the inlet is fixed to 25\degr. A power of 6~W is dissipated on the silicon sensor surface immediately above the micro-channel cooling manifold.

	\begin{table}[ht!]
	\caption{Physical and chemical properties of water and silicon used in the simulation: the density $\rho$, the constant pressure heat capacity $C_p$, the initial temperature $T$ and the thermal conductivity $\kappa$.}
	\label{tab:water}
        \begin{center}
	\begin{tabular}{lcc}
		\hline
       		& Water		& Silicon \\ \hline	
	$\rho$ 	& 997 		& 2329	\\ 
	$C_p$ 	[J kg$^{-1}$K$^{-1}$] & 4181.3	& 713	\\ 
     T [K] 	& 293		& 293		\\ 
     $\kappa$ [W m$^{-1}$K$^{-1}$]	& 0.58 		& 120		\\ \hline

	\end{tabular}
	\end{center}
	\end{table}


Results from a number of simulations with varying liquid flow are presented in Table~\ref{tab:Q_water_sim}. The liquid temperature at the outlet runs several degrees hotter than the inlet side. This reflects the heating of the liquid as it progresses through the manifold. The fourth column in Table~\ref{tab:Q_water_sim} presents the simulated outlet temperatures. From these results one can calculate the heat absorbed by the cooling liquid:
\begin{equation}
	\dot{Q_a}  = \dot{m}·C_p\Delta T,
	\label{eq:CFX}
\end{equation}
where $\dot{Q}$ is the absorbed power, $\dot{m}$ the liquid mass flow, $C_p$ is the constant pressure heat capacity of the liquid (the value for water is given in table \ref{tab:water}) and $\Delta T$ is the water temperature difference between the inlet and the outlet. The efficiency $\varepsilon$ of the simulated heat exchanger is obtained by comparing the absorbed power $Q_a$ with the power absorbed by the liquid flow.

	\begin{table*}[ht!]
        \begin{center}
	\begin{tabular}{rcccccc}
		\hline
		Dissipated 	& Liquid mass 	& Inlet temp.	& Outlet temp.	& Absorbed	& efficiency	\\ 
    	power $Q_d$ [W]	& flow $\dot{m}$	[kg/s] 	& T$_{in}$ [K]	& T$_{out}$ [K]	& power $Q_{a}$ [W]	& $\varepsilon$ [\%]	\\ \hline
		5.82				&6.98$\times10^{-5}$& 298			& 315.1			& 5.01			&  86			\\ 
		5.99				&9.70$\times10^{-5}$& 298			& 312.0			& 5.69			&  95			\\ 
		6.12				&1.29$\times10^{-4}$& 298			& 303.9			& 5.90			&  96			\\ 
		6.17				&1.68$\times10^{-4}$& 298			& 306.5			& 5.96			&  97			\\ 
		6.19				&2.06$\times10^{-4}$& 298			& 305.0			& 6.01			&  97			\\ 
		6.21				&2.42$\times10^{-4}$& 298			& 304.0			& 6.07			&  98			\\ 
		6.23				&2.72$\times10^{-4}$& 298			& 303.4			& 6.10			&  98			\\ 
		6.25				&3.05$\times10^{-4}$& 298			& 302.8			& 6.14			&  98			\\ 
		6.25				&3.45$\times10^{-4}$& 298			& 302.3			& 6.16			&  98			\\ 
		6.25				&3.83$\times10^{-4}$& 298			& 301.9			& 6.18			&  99			\\ 
		 
	\end{tabular}
	\end{center}
	\caption{The results of a number of simulations with different mass flow of cooling liquid. The first five parameters appear in Eq.~\ref{eq:CFX}. The efficiency $\varepsilon$ relates the heat $\dot{Q}$ absorbed by the coolant to the total applied power. }
	\label{tab:Q_water_sim}
	\end{table*}

\begin{figure}[h!]
\centering
\includegraphics[width=0.9\columnwidth]{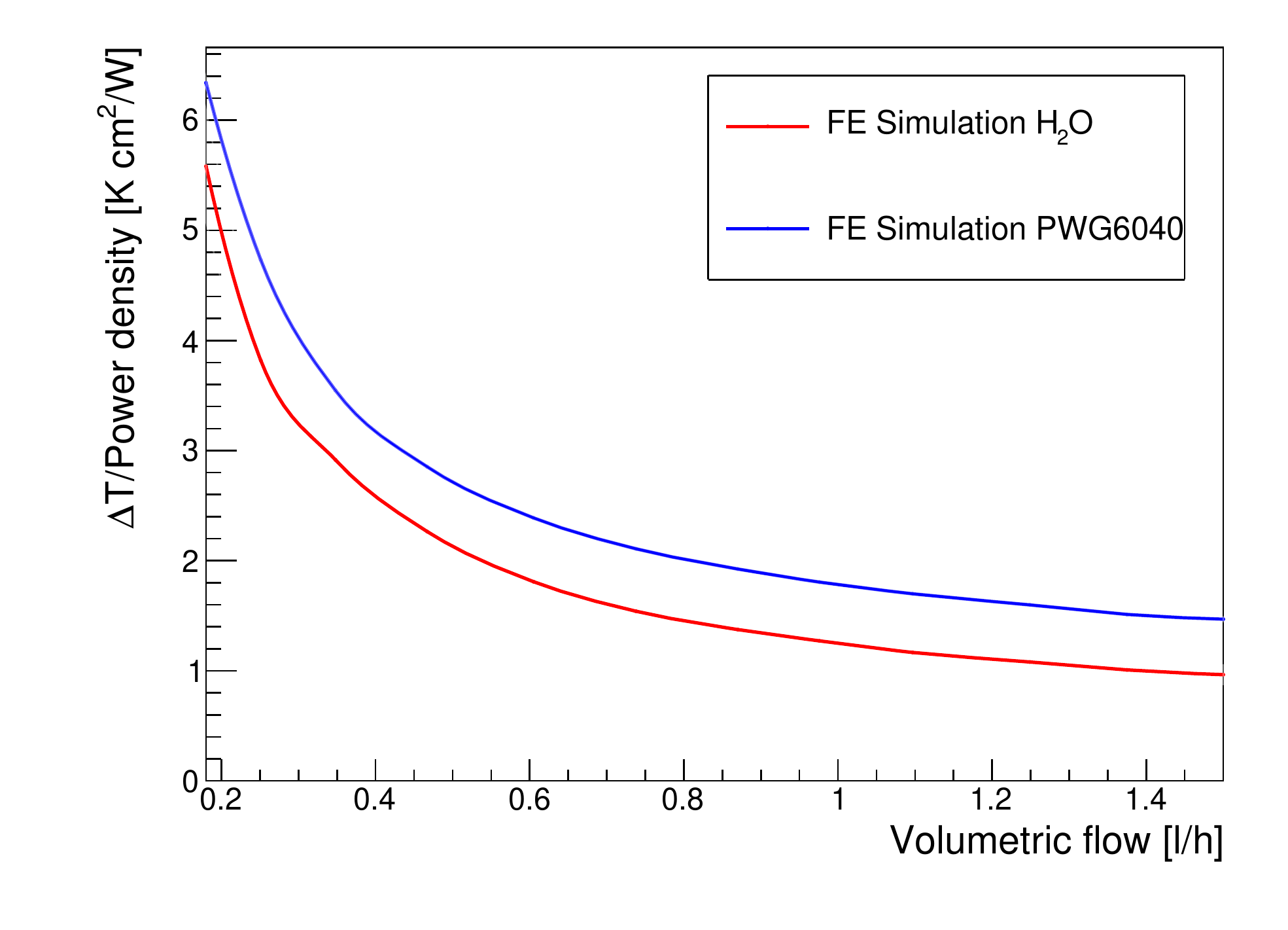}
\caption{Finite element simulation results for the temperature gradient per unit power density as function of the volumetric flow of the coolant. The lower, red curve represent the situation of the laboratory measurement in Section~\ref{sec:results}, where water is used as a coolant. The upper, blue curve is obtained when the water is replaced with an alcohol-water mixture (PWG6040).}
\label{fig:coolant}
\end{figure}

We can now use the simulation to rapidly explore the thermal performance of different designs under conditions that are hard to mimic in the laboratory. As an example we evaluate the effect on the cooling performance of the choice of the coolant. In Fig.~\ref{fig:coolant} the figure-of-merit of several choices are compared. The lower, red curve in Fig.~\ref{fig:coolant} represent the situation of the laboratory measurement in Section~\ref{sec:results}, where water is used as a coolant. The upper, blue curve in the same figure is obtained when the water is replaced with an alcohol-water mixture (PWG6040). 

The cooling performance is somewhat degraded for most other choices for the coolant. The temperature gradient found in the simulation is in approximate 
agreement with a naive scaling by the ratio of heat capacities.

\section{Test set-up}
\label{sec:laboratory}
In this Section a setup is described to measure the cooling performance of the silicon sensor with embedded micro-channels under a realistic heat load. The aim is to measure the temperature of the sensor surface as a function of the operational parameters, primarily the power consumption and liquid flow. The setup also monitors the position of the sensor to sub-$\mu\mathrm{m}$ precision.

\begin{figure*}[t!]
\centering
\includegraphics[width=0.89\textwidth]{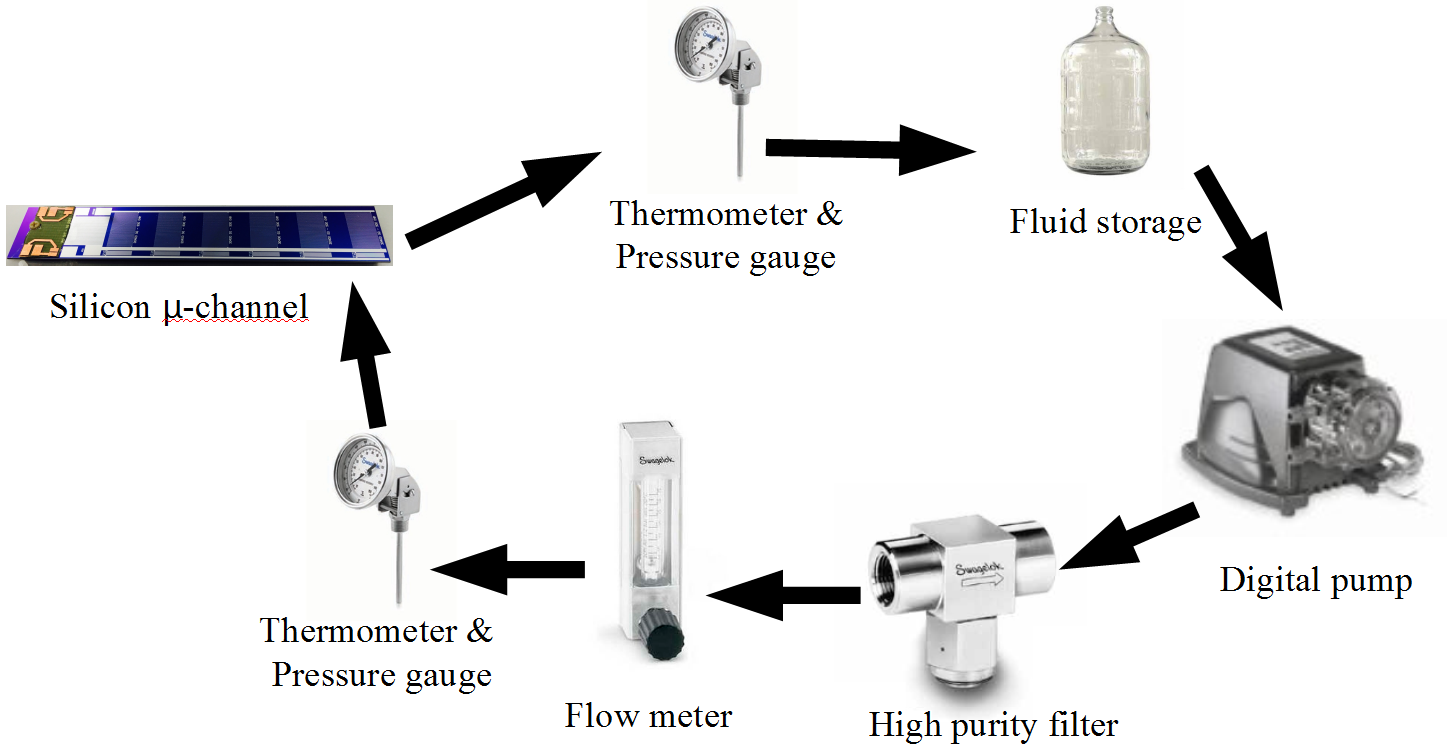}
\caption{On the left, schematic layout of the laboratory set up and on the right real setup ready for measurement, with tube connexions and sensors placed.}
\label{fig:layout}
\end{figure*}

The measurements are performed deionized water as the cooling fluid. The measurements are representative for any monophase cooling system. In Section~\ref{sec:simulation} we extrapolate to other cooling liquids using the Finite Element simulation. The components to control and monitor the inputs, water flow, pressure and temperature at the inlet and outlet, are presented in Fig.~\ref{fig:layout}.


The water flow is established using a digital, variable-speed peristaltic pump (Stenner, SVP1). The flow can be regulated between a fraction of a liter per hour and several liters per hour. The maximum pressure the system can supply is 6 bars. To assure that there are no microparticles which obstruct the microchannels in the silicon ladder, a filter with a maximum particle size of 15~$\mu\mathrm{m}$ is inserted in the circuit. The liquid flow is measured using a flow meter and temperature sensors and pressure gauges are placed immediately before the inlet and after the outlet of the silicon ladder. All devices use standard SwageLock connectors. The only custom piece is the connector to the micro-channel circuit described in Section~\ref{sec:connectors}.

The temperature of the sensor is measured using a thermal infra-red camera (FLIR Systems ThermaCAM SC500) with a field of view that includes the entire surface of the sensor. The temperature readings are calibrated to the readings of sensors in direct contact with the silicon (Pt1000). The contact sensors are only used during calibration runs, so as to avoid any mechanical disturbance. The temperature calibration is applied as a global correction for the average emissitivity of the materials (silicon and aluminium). A more precise reading is obtained with a second single-spot infra-red sensor (Optris CTLaser OPTCTLLTFCF1). This sensor is used to measure the temperature of the hottest point of the silicon ladder. 
	
The mechanical impact of the micro-channel cooling is determined using a triangulating sensor (Micro-Epsilon OptoNCDT ILD2300-10) located above the sensor. This sensor provides readings of the distance to the ladder perpendicular to the sensor surface with a resolution of 0.15 $\mu$m. The read-out frequency of 1 kHz is sufficient to monitor vibrations in the most important frequency range. To provide maximum sensitivity to deformations and vibrations, the sensor is supported only on one side, by the connector described in Section~\ref{sec:connectors}. The distance sensor is placed close to the edge furthest away from the support, for the same reason. The displacement sensor, sensor and support structure are placed on an optical table with pneumatic damping to isolate the system from external vibrations.

\section{Thermal performance}
\label{sec:results}

The cooling capacity of the micro-manifold integrated in the silicon is characterized in a series of measurements of the maximum temperature on the sensor surface. The measurement is repeated for a number of different settings of the pump speed, corresponding to a liquid flow ranging from a fraction of a liter/hour to 1.5~l/h, and for different values of the power dissipated in the end-of-ladder area. For all measurement the outlet of the cooling circuit is at atmospheric pressure and the ambient temperature between 23-25$\degr\mathrm{C}$.
	
\begin{figure}[h!]
\centering
\includegraphics[width=0.9\columnwidth]{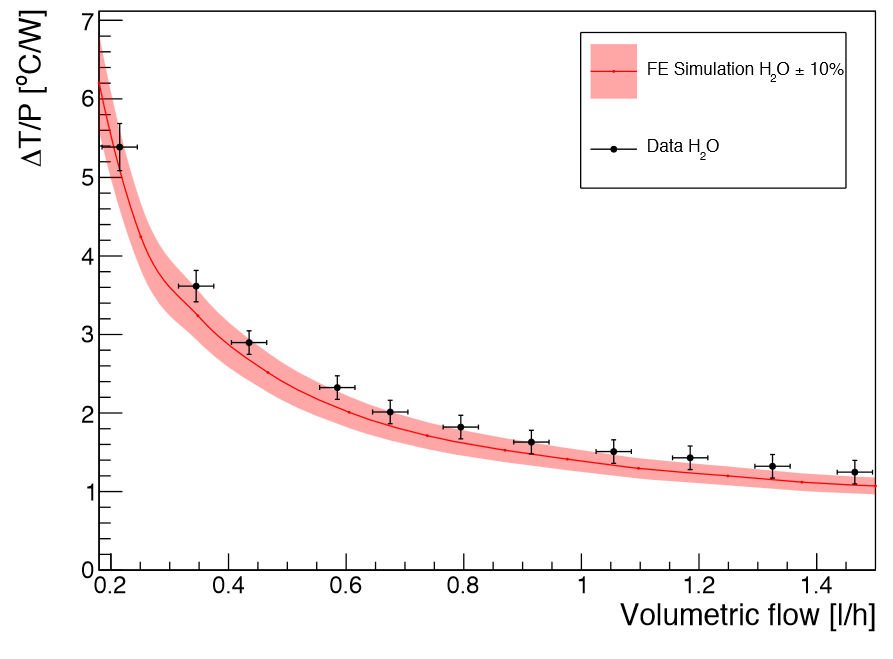}
\caption{Temperature gradient of the hottest point on the sensor surface divided by the dissipated power as a function of the water flow.}
\label{fig:T-Q}
\end{figure}

	In Figure \ref{fig:T-Q} the temperature gradient from the hottest point on the sensor to the coolant at the inlet of the micro-channel circuit is plotted as a function of liquid flow. The temperature difference is divided by the dissipated power because power is not exactly constant (as the resistance of the heaters on the silicon sensor depends on the temperature). 
	
\begin{figure}[h!]
\centering
\includegraphics[width=0.9\columnwidth]{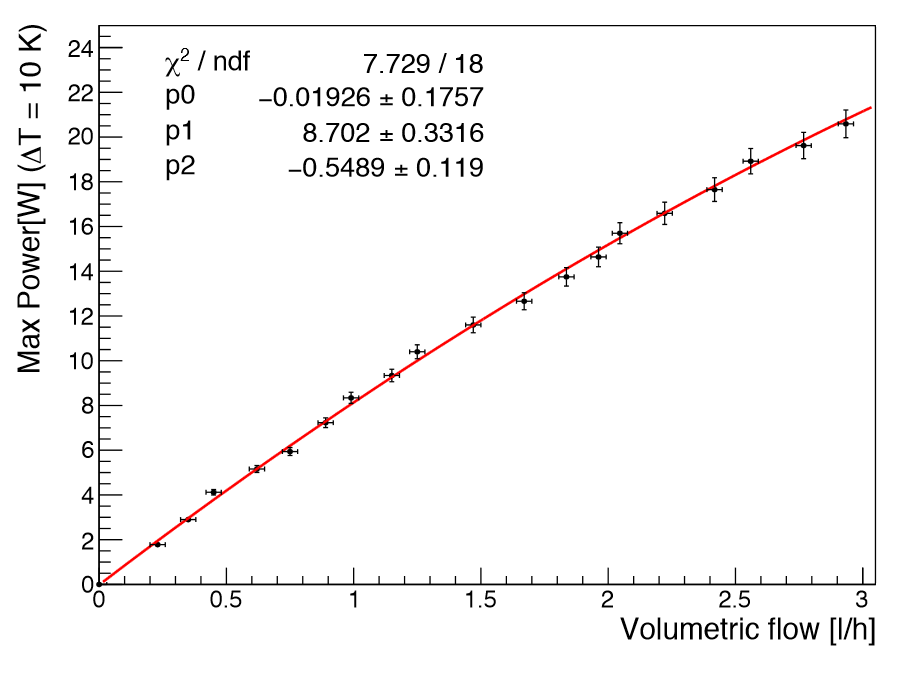}

\caption{Dissipated power on the silicon ladder that yields a thermal gradient of  10$^{\circ}\mathrm{C}$ for different liquid flow rates. }
\label{fig:P-Q}	
\end{figure}

With a series of measurement at different power the maximum power for a given temperature gradient can be determined. An example is shown in Figure \ref{fig:P-Q}, where a maximum gradient of 10$^{\circ}\mathrm{C}$ is allowed. At a liquid flow of 3~l/h, the integrated micro-channel circuit removes up to 20~W/${cm^2}$ while maintaining the temperature gradient below 10$^{\circ}\mathrm{C}$. The curve is nearly linear over the interval of the measurements. While some sign of saturation is visible for the maximum flow rate, the derivative is still two thirds of that at the origin. This implies that the system is still not at its limit: a further increase of the liquid flow can provide further cooling power.

The pressure drop over the micro-channel cooling circuit is 1.5 bar when water flows at 3~l/h. The pressure drop is negligible (below 0.1 bar) at the smallest flow rate of 0.2~l/h.

\section{Mechanical impact}
\label{sec:results3}

\begin{figure*}[t!]
\centering
\includegraphics[width=0.7\linewidth]{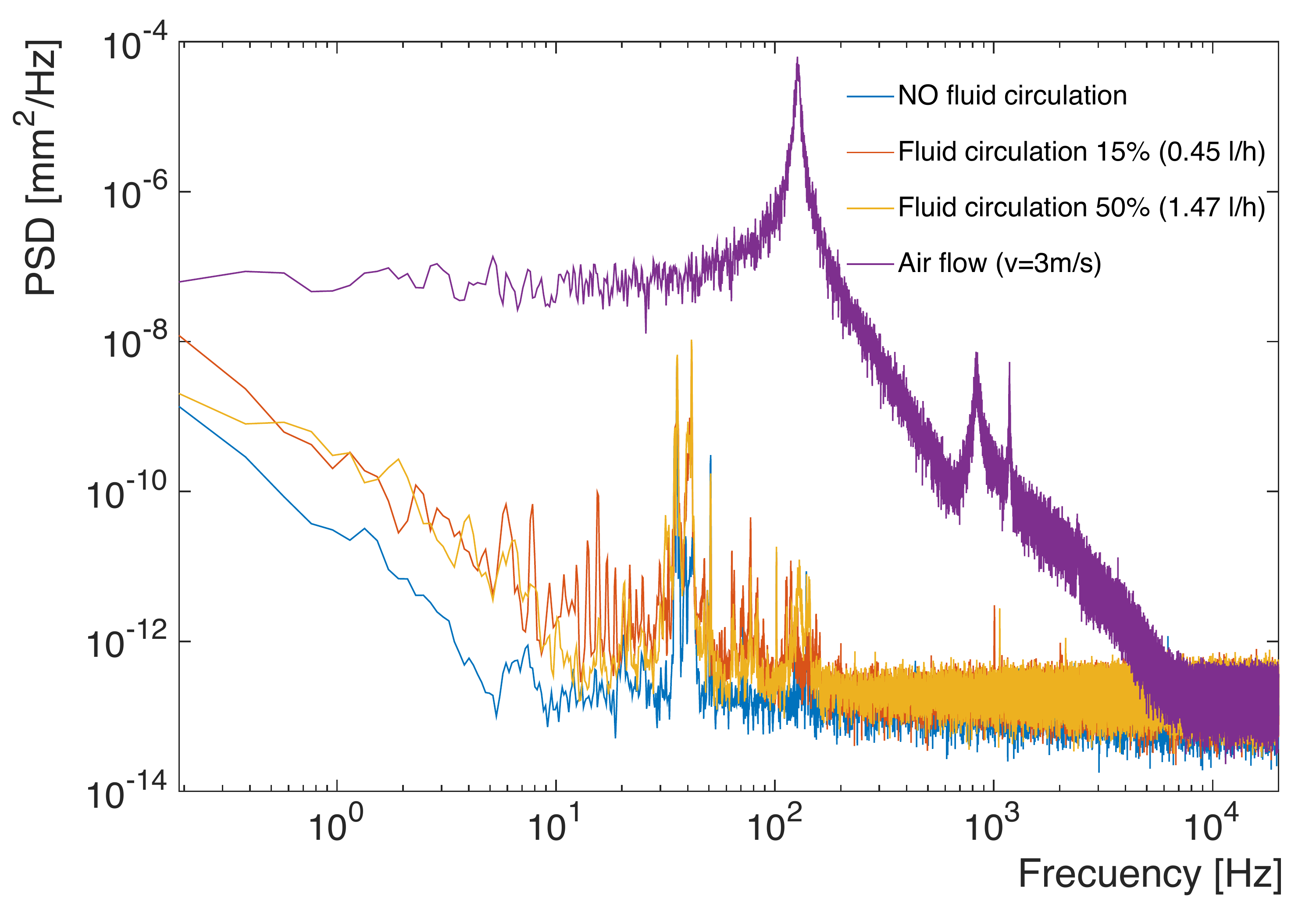}
\caption{The spectral power density of four time series registered with the triangulating sensor: without any cooling (light blue), with a liquid flow of 0.45~l/h (red) and 1.47~l/h (yellow) and with an air flow of 3~m/s (purple). }
\label{fig:frecuencies}
\end{figure*}

The mechanical stability of the sensor is monitored throughout the measurements. To magnify the deformations the sensor is only supported on one end and the triangulating sensor is located on the opposite end of the ladder. The sensor measures the distance to the ladder. The measurement is therefore sensitive to deformations and vibrations in the direction perpendicular to the sensor area.

Several time series of the distance measurement are recorded under different conditions. The vibrations introduced by the background are measured without liquid flow. The peak-to-peak interval of all distance measurements for this null experiment is less than 0.7~$\mu\mathrm{m}$. The Root Mean Square (RMS) of the readings is 0.3$\mu\mathrm{m}$. When the measurement is repeated for the maximum liquid flow of 1.47 $\mathrm{l}/\mathrm{h}$ the excursions increases only very slightly: the peak-to-peak interval is approximately 1~$\mu\mathrm{m}$, the RMS increases to 0.4$\mu\mathrm{m}$. To check the sensitivity of the setup, the measurements were repeated with an air flow of 3~m/s. This induces a very significant vibrations, with a peak-to-peak interval of 130$\mu\mathrm{m}$, demonstrating the sensitivity of the setup to vibrations.


The readings in the time domain are translated to the frequency domain using
a fast Fourier transform. The spectral power
density if plotted in Fig.~\ref{fig:frecuencies} for frequencies ranging from
a fraction of a~Hz to several~kHz. The eigenfrequency of the ladder (at 150~Hz)
is clearly visible in the spectrum corresponding to the air flow. The other 
spectra are quite close to the null spectrum, without air or liquid flow.
The liquid flow seems to induce very subtle vibrations, with sub-$\mu{m}$ 
magnitude and frequencies in the 1-100~Hz domain.



\section{Conclusions}
\label{sec:conclusions}

In this paper we have proposed a process to monolithically integrate a 
micro-channel cooling circuit in position-sensitive silicon sensors.
The process allows to combine standard double-sided processing of the sensor
with the precise etching of the cooling manifold and a flexible thinning 
procedure, while minimizing the number of processing steps with delicate 
handling of thin structures. 

We aim for a low-pressure (several bars), mono-phase cooling solution, 
with a limited liquid
flow, which is adequate in room-temperature applications with modest power
consumption. A good example is the tracker and vertex detector of a 
future energy-frontier $e^+e^-$ collider, but this solution is expected
to be applicable in a broad range of application.

A small number of prototype structures has been produced with a channel cross section 
of 100 $\times$ 380 $\mu\mathrm{m}^2$. 
The inlet and outlet with an approximately rectangular cross section 
of 340 $\times$ 380 $\mu\mathrm{m}$ are accessible from the sensor edge at 
the narrow end of the silicon ladder.
The micro-channel circuit is connected to the main cooling circuit through
custom 3D-printed connectors that interface the narrow inlet and outlet
to commercial high-pressure connectors. The connection is sealed using
Araldite 2011 glue. For the present purpose these
plastic structures are found to be sufficiently leak tight. We find they
resist a pressure of 180 bars, which is well in excess of the requirement.
A future evolution of these connectors could provide a low-Z, low-mass
solution in the tracking volume of the experiment.

The cooling performance is measured by applying a power of up to 10~$W$
on the surface of the silicon sensor immediately above the integrated
cooling circuit. The absence of thermal barriers leads to very efficient
evacuation of the dissipated heat. The temperature gradient from the hottest
point on the sensor to the liquid can be kept below 2~K for
a power density of 6~$W/cm^2$ and a liquid flow of 1 $l/h$, corresponding
to a thermal figure of merit of close to 1.

A finite-element simulation of the cooling circuit is able to reproduce the 
observed cooling performance, predicting the temperature gradient 
as a function of liquid flow and power dissipation to within 10\%.
The simulation is used to simulate a situation, where the power is dissipated 
on an ASIC that is connected to the sensor through bump bonds.
This leads to an increase of the temperature gradient from heat 
source to heat sink of 5~K.

We find the liquid flow has no significant impact on the mechanical stability
of the sensor: the amplitude of vibrations is measured to be 
smaller than 1$\mu\mathrm{m}$, compatible with the null result
when no liquid circulates.

\section*{Acknowledgement} 
Part of the funding for the studies reported in this paper 
stems from the European Union’s Horizon 2020 Research and Innovation programme under Grant Agreement no. 654168. We thank our partners in the AIDA2020 thermo-mechanical work package, and in particular the CERN group (P. Petagna, A. Mapelli, J. Noel) for access to the leak and pressure test.

\bibliographystyle{JHEP}
\bibliography{depfet}{}

\providecommand{\href}[2]{#2}\begingroup\raggedright\begin{thebibliography}{10}

\bibitem{bib:ilddbd}
{\bf ILC community} Collaboration, {\it {The International Linear Collider,
  Technical Design Rep ort}}, .

\bibitem{Baer:2013cma}
H.~Baer, T.~Barklow, K.~Fujii, Y.~Gao, A.~Hoang {\em et.~al.}, {\it {The
  International Linear Collider Technical Design Report - Volume 2: Physics}},
  \href{http://arXiv.org/abs/1306.6352}{{\tt 1306.6352}}.

\bibitem{Behnke:2013lya}
T.~Behnke, J.~E. Brau, P.~N. Burrows, J.~Fuster, M.~Peskin {\em et.~al.}, {\it
  {The International Linear Collider Technical Design Report - Volume 4:
  Detectors}},  \href{http://arXiv.org/abs/1306.6329}{{\tt 1306.6329}}.

\bibitem{Linssen:2012hp}
L.~Linssen, A.~Miyamoto, M.~Stanitzki and H.~W~eerts, {\it {Physics and
  Detectors at CLIC: CLIC Conceptual Design Report}},
  \href{http://arXiv.org/abs/1202.5940}{{\tt 1202.5940}}.

\bibitem{Gomez-Ceballos:2013zzn}
M.~Bicer, H.~Duran~Yildiz, I.~Yildiz, G.~Coignet, M.~Delmastro {\em et.~al.},
  {\it {First Look at the Physics Case of TLEP}},
  \href{http://arXiv.org/abs/1308.6176}{{\tt 1308.6176}}.

\bibitem{Mapelli:2012zz}
A.~Mapelli, P.~Petagna, K.~Howell, G.~Nuessle and P.~Renaud, {\it {Microfluidic
  cooling for detectors and electronics}},  {\em JINST} {\bf 7} (2012) C01111.

\bibitem{Abelevetal:2014dna}
{\bf ALICE} Collaboration, B.~Abelev {\em et.~al.}, {\it {Technical Design
  Report for the Upgrade of the ALICE Inner Tracking System}},  {\em J. Phys.}
  {\bf G41} (2014) 087002.

\bibitem{Buytaert:2013hka}
J.~Buytaert {\em et.~al.}, {\it {Micro channel evaporative $CO_2$ cooling for
  the upgrade of the LHCb vertex detector}},  {\em Nucl. Instrum. Meth.} {\bf
  A731} (2013) 189--193.

\bibitem{Nussle:2014sza}
G.~N{\"u}ssle, A.~Mapelli, M.~Morel, P.~Petagna, G.~Romagnoli and K.~Howell,
  {\it {NA62 GigaTracKer Cooling with Silicon Micro Channels}},  in {\em
  {Proceedings, 14th ICATPP Conference on Astroparticle, Particle, Space
  Physics and Detectors for Physics Applications (ICATPP 2013)}}, pp.~525--530,
  2014.

\bibitem{Abe:2010gxa}
{\bf Belle-II} Collaboration, T.~Abe {\em et.~al.}, {\it {Belle II Technical
  Design Report}},  \href{http://arXiv.org/abs/1011.0352}{{\tt 1011.0352}}.

\bibitem{Alonso:2012ss}
{\bf DEPFET} Collaboration, O.~Alonso {\em et.~al.}, {\it {DEPFET active pixel
  detectors for a future linear $e^+e^-$ collider}},
  \href{http://arXiv.org/abs/1212.2160}{{\tt 1212.2160}}.

\bibitem{VTXdepfetmech2}
L.~Andricek, G.~Lutz, R.~H. Richter and M.~Reiche, {\it {Processing of
  ultra-thin silicon sensors for future e+ e- linear collider experiments}},
  {\em IEEE Trans.Nucl.Sci.} {\bf 51} (2004) 1117--1120.

\end{thebibliography}\endgroup
\end{document}